\documentclass[aps, prb,  twocolumn, notitlepage,superscriptaddress]{revtex4-1}
\usepackage{newtxtext, newtxmath}
\usepackage{hyperref}
\hypersetup{
  colorlinks   = true, 
  urlcolor     = blue, 
  linkcolor    = blue, 
  citecolor   = red 
}

\usepackage{graphicx}
\usepackage{ulem}
\usepackage{xcolor}
\usepackage{physics}
\usepackage{booktabs}
\usepackage{multirow}
\usepackage{numprint}
\setcitestyle{numbers,square}

\renewcommand{\vec}[1]{\mathbf{#1}}

\newcommand{\mat}[1]{\bar{\bar{\mathbf{#1}}}}
\newcommand{\msigma}{\bar{\bar{\sigma}}}

\newcommand{\gr}{\mat{G}}
\newcommand{\vk}{\vec{k}}

\newcommand{\re}{\vec{R}_\mathrm{e}}
\newcommand{\rp}{\vec{R}_\mathrm{P}}

\newcommand{\tauu}{\tau_u}
\newcommand{\ri}{\vec{R}_i}
\newcommand{\rj}{\vec{R}_j}
\newcommand{\Ri}{\vec{R}_{I}}
\newcommand{\Rj}{\vec{R}_{J}}
\newcommand{\Rij}{\vec{R}_{IJ}}

\newcommand{\qs}{\vec{q}_\mathrm{S}}
\newcommand{\qp}{\vec{q}_\mathrm{P}}

\newcommand{\dqpv}{\delta_{\vec{\qp} \mathrm{\nu}}}

\newcommand{\Si}{\vec{S}_{I}}
\newcommand{\Sj}{\vec{S}_{J}}
\newcommand{\Jani}{\mat{J}^\mathrm{ani}}
\newcommand{\Dij}{\vec{D}_{IJ}}

\begin{document}

\title{First--Principles Spin--Lattice Coupling from Downfolded Electron--Phonon Interaction}

\author{Xu He}
\affiliation{Physique Th\'eorique des Mat\'eriaux, Q-Mat, Universit\'e de
  Li\`ege, B-4000 Sart-Tilman, Belgium}
\email{x.he@uliege.be}

\author{\'Alvaro Adri\'an Carrasco \'Alvarez}
\affiliation{European Theoretical Spectroscopy Facility, Institute of Condensed Matter and Nanosciences, Université catholique de Louvain, Chemin des Étoiles 8, B-1348 Louvain-la-Neuve, Belgium}

\author{Gian-Marco Rignanese}
\affiliation{European Theoretical Spectroscopy Facility, Institute of Condensed Matter and Nanosciences, Université catholique de Louvain, Chemin des Étoiles 8, B-1348 Louvain-la-Neuve, Belgium}
\affiliation{WEL Research Institute, avenue Pasteur 6, 1300 Wavre, Belgium.}

\author{Eric Bousquet}
\affiliation{Physique Th\'eorique des Mat\'eriaux, Q-Mat, Universit\'e de Li\`ege, B-4000 Sart-Tilman, Belgium}

\author{Samuel Poncé}
\affiliation{European Theoretical Spectroscopy Facility, Institute of Condensed Matter and Nanosciences, Université catholique de Louvain, Chemin des Étoiles 8, B-1348 Louvain-la-Neuve, Belgium}
\affiliation{WEL Research Institute, avenue Pasteur 6, 1300 Wavre, Belgium.}

\author{Matthieu J. Verstraete}
\affiliation{Nanomat, Q-Mat, and European Theoretical Spectroscopy Facility, Universit\'e de Li\`ege, B-4000 Li\`ege, Belgium}
\affiliation{ITP, Physics Department, Utrecht University 3508 TA Utrecht, The Netherlands}

\begin{abstract}
  We present a method to calculate spin--phonon coupling parameters from
  first-principles perturbation theory by downfolding the
  electron--phonon coupling (EPC). We exploit the localized nature of
  magnetic moments and atomic displacements by working in the Wannier representation of
  the electronic Hamiltonian and the EPC matrix. The spin system is mapped to a classical Heisenberg
 Hamiltonian, whose parameters are obtained by treating local
 spin rotations as a perturbation within a Green's-function formalism. The spin and phonon
 perturbations are connected through the EPC parameters, which enter as
 lattice-induced perturbations to the tight-binding Hamiltonian. By combining
 these lattice perturbations with local spin rotations, we obtain real-space
 derivatives of magnetic exchange parameters without performing displaced
 magnetic supercell calculations. We illustrate the method on SrMnO$_3$ and
 show that it can be integrated directly into standard workflows.

\end{abstract}

\maketitle
\section{Introduction}
The Heisenberg model and its associated spin dynamics have long been fundamental tools for understanding the magnetic properties of materials, enabling researchers to bridge the gap between microscopic electronic interactions and macroscopic magnetic behavior. However, as the field of materials science pushes toward the limits of efficiency and speed, the precise treatment of the coupling between spin and lattice degrees of freedom has become increasingly critical. This spin-phonon coupling is not merely a correction to the magnetic Hamiltonian but is the physical origin of many interesting phenomena, including strong modulation of magnon and phonon lifetimes~\cite{PhysRevB.94.014431}, the
emergence of new excitations~\cite{PhysRevLett.83.2062} and
magnon-to-phonon~\cite{maehrlein2018dissecting,PhysRevB.100.014430}  conversion, which could have a
strong impact on spintronics, ultrafast demagnetization, magnetocalorics,
spincaloritronics, and multiferroics~\cite{wang2012atomistic}.

Traditionally, the dynamics of magnetic structures are described using the Heisenberg Hamiltonian, which captures the quadratic interactions between localized spins. While such a model provides a robust framework for simulating magnetic properties via spin dynamics, it typically assumes a rigid lattice where atomic positions remain fixed. In real materials, thermal and quantum fluctuations of the atomic nuclei excite phonons which dynamically modulate the crystal field and intersite electronic hopping. This modulation in turn alters the exchange interactions between the spins, giving rise to an intrinsic coupling between the magnetic and lattice degrees of freedom~\cite{waller1932magnetization,vleck1940paramagnetic}.

Several related terms are used to describe the coupling between spin and lattice
degrees of freedom, including spin--lattice coupling (SLC), spin--phonon
coupling (SPC), magnon--phonon coupling (MPC), and magnon--lattice coupling
(MLC). These terms emphasize different representations: one may start from
localized spin rotations or from collective magnon excitations, and similarly
from local atomic displacements or collective phonons. The corresponding
parameters can often be transformed into one another, sometimes with additional
approximations. For example, the magnon spectrum can be obtained from a spin
Heisenberg model through spin dynamics or linear spin-wave theory, while the
phonon spectrum can be obtained from a lattice model through molecular dynamics
or the harmonic approximation. Although these terms are often used
interchangeably in the literature, they are not strictly identical. In this
work, we use the term SLC because our formulation is based on localized
real-space spin and lattice perturbations.

The spin-lattice coupled dynamics has been formulated in several works~\cite{PhysRevB.78.024434, tranchida2018massively,
  PhysRevB.99.104302, PhysRevLett.86.898,ma2016spilady,PhysRevMaterials.1.074404,gonze2019abinit,slcmultibinit,mankovsky2023spin}, and its application to specific materials~\cite{PhysRevB.78.024434,PhysRevB.86.214423,PhysRevLett.113.165503} has shown that the quantitative analysis of some phenomena necessitates the inclusion of both spin and lattice degrees of freedom.
The total Hamiltonian can be written as the sum of the spin, lattice and
coupling terms: $H= H^\mathrm{spin} + H^\mathrm{latt}+ H^\mathrm{SLC}$.

Both the lattice and the spin parts of the Hamiltonian can be constructed from the results of first-principles calculations. The spin part can be written
in a Heisenberg-like form: $H^\mathrm{spin} = -\sum_{IJ} ( J^\mathrm{iso}_{IJ}\Si\cdot\Sj + \Si \Jani_{{IJ}} \Sj + \Dij\cdot (\Si\cross\Sj) ) $,
considering the isotropic exchange interaction, where $I$ and $J$ denote the
indices of the atoms. With this sign convention, a positive $J_{IJ}^\mathrm{iso}$
favors ferromagnetic alignment and a negative $J_{IJ}^\mathrm{iso}$ favors antiferromagnetic
alignment. Other terms such as
the Dzyaloshinskii-Moriya interaction $\Dij$ (DMI), the anisotropic exchange $\Jani$, the single ion
anisotropy, or  the magnetic dipole-dipole interaction can also be included, and they would also be coupled with the lattice displacements. In this work, we focus on the coupling between the
isotropic exchange interaction and lattice distortions, which to first order is
parametrized with:
\begin{equation}
  \label{eq:OijuTijuv}
 O_{IJu}= -\frac{d^3E }{d\Si d\Sj d\tauu}= \frac{dJ_{IJ}^\mathrm{iso}}{d\tauu} \,,
\end{equation}
where $\tauu$ is the displacement of an atom along one direction, and $u$
contains the atom and direction indices.
Higher-order polynomial couplings can be generalized but are more complex to compute from first principles.

 Many methods can be used to model the lattice vibrations,
 from those based on the simple harmonic approximation to more sophisticated
 models which include anharmonic interactions~\cite{wojdel2013, secondprinciple}, and more recently machine learning interatomic potentials (MLIPs)~\cite{mishin2021machine}.
 The harmonic phonons can be
 modeled with interatomic force constants, which can be calculated with either
 density functional perturbation theory (DFPT)~\cite{baroniRMP, abinitref}, or finite differences (FD)~\cite{phonopy}
 from first-principles.
 
In the case of the spin part of the Hamiltonian, there are several methods for computing the parameters in spin models. One
of the most commonly used methods is based on
first-principles total-energy calculations and energy mapping~\cite{mappingparameter}.
Density Functional Perturbation Theory with respect to external magnetic fields was developed
by Savrasov~\cite{Savrasov1998_prl_81_2570}.
Alternatively, the Generalized Bloch Theorem (GBT)~\cite{herring1966magnetism, sandratskii1986} can
be used to calculate the total energy of different spin-spiral states, from which
the exchange parameters can be extracted. 
Finally, the magnetic force theorem (MFT) method makes use of Green's functions,
taking the local spin rotation as a perturbation, as proposed in the Liechtenstein, Katsnelson, Antropov and Gubanov formalism (LKAG)
\cite{LKAG}. The MFT method has been widely used thanks to its simplicity (only unperturbed ground state calculations are required as input), efficiency (the Green's function expressions are direct and closed-form expressions), and accuracy (for many purposes the perturbative approximation is sufficient).  This method was extended to deal with
correlated materials~\cite{KatsnelsonLKAGcorrelated},
magneto-crystalline-anisotropy~\cite{LKAGMAE, LKAGfulltensor},
DMI~\cite{LKAGDMI0,LKAGDMI1, LKAGSIADMI, LKAGfulltensor} in fully
relativistic cases with non-collinear spin ground states~\cite{LKAGrelativistic}, and higher-order magnetic interactions~\cite{LKAGanharmonic, LKAGanharmonic2}


In the literature, several methods for estimating SLC parameters have been developed, including supercell-based finite-difference schemes~\cite{sadhukhan2022spin, PhysRevB.99.104302, mappingparameter, fang2025efficient} and response-function approaches such as Berry-curvature calculations~\cite{ren2024adiabatic}, two-particle Bethe--Salpeter equation~\cite{le2025magnon}, or time-dependent DFPT~\cite{delugas2023magnon, gorni2023first}. While these methods provide comprehensive insights, they typically involve heavy supercell calculations or complex dynamical spectral frameworks.

We propose instead a perturbative downfolding approach that extracts real-space SLC parameters directly from the primitive-cell electron--phonon coupling (EPC). By extending the MFT to include atomic displacements as a simultaneous perturbation, our method computes derivatives of exchange parameters from Wannier-interpolated EPC and avoids the need for displaced magnetic supercells. The method provides an efficient and orbital-resolved framework that is closely related to the work of Mankovsky \textit{et al.}~\cite{mankovsky2023spin}. This approach bridges the gap between simple force fitting and advanced response theories, offering a practical way to simulate complex spin-lattice dynamics.

\begin{figure}[htbp]
  \centering
  \includegraphics[width=\linewidth]{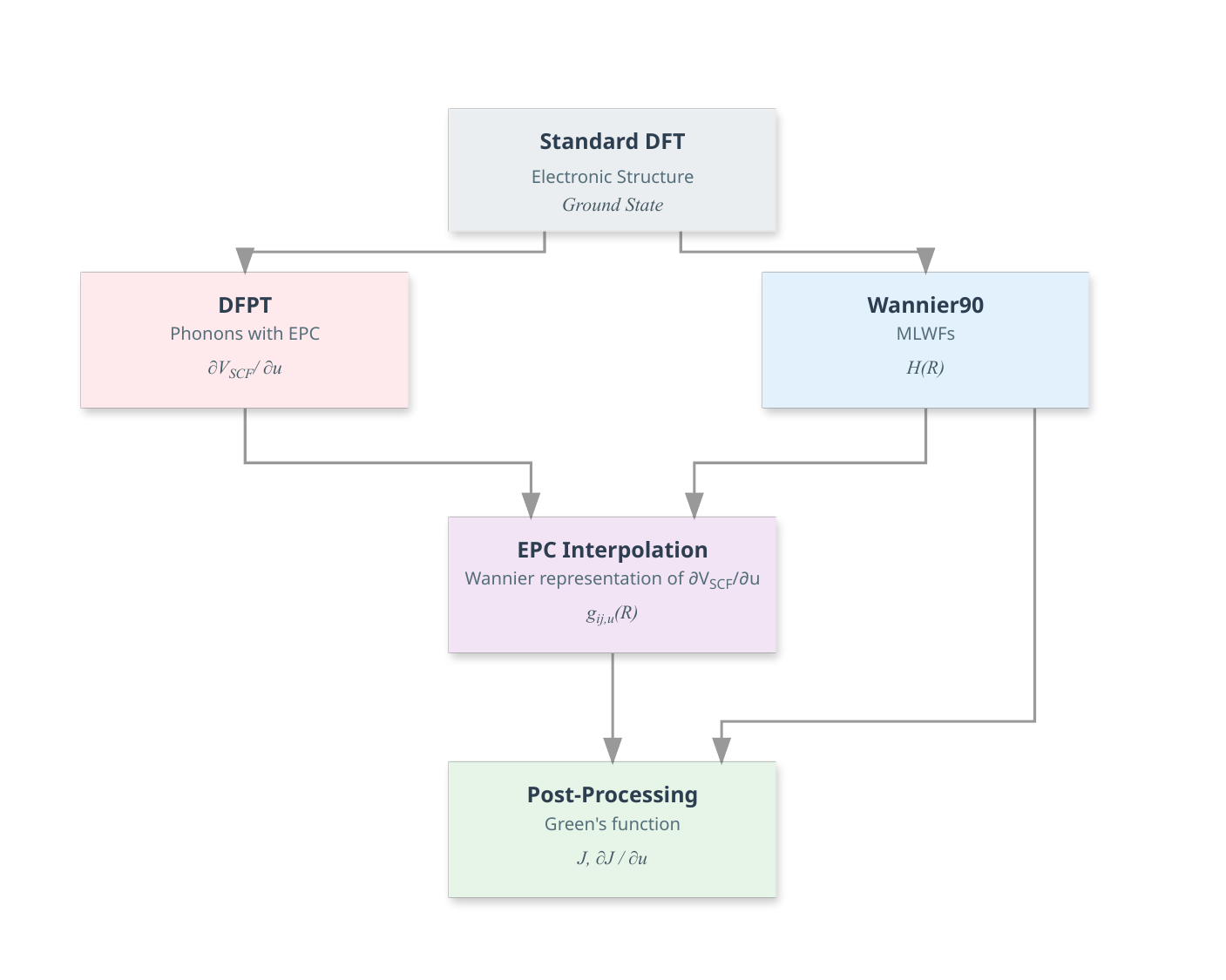}
  \caption{Schematic workflow of the Wannier-based perturbative method. A standard DFT calculation provides the reference electronic structure, DFPT yields phonons and the variation of the self-consistent potential, Wannier functions and the real-space Hamiltonian can be constructed  with \texttt{Wannier90}~\cite{mostofi2008wannier90}, EPC in the Wannier basis from the DFPT result can be obtained, for example, with \texttt{EPW}~\cite{ponce2016epw,carrascoalvarez2025}, and the magnetic Green's functions used to evaluate the SLC parameters are built with \texttt{TB2J}~\cite{TB2J}.}
  \label{fig:flow}
\end{figure}

We describe the MFT method for the
calculation of the spin interaction parameters, and the Wannier representation of
the EPC, in Section~\ref{sect:mixing_SLC}.
In Section~\ref{sect:example} we demonstrate the method with the example of SrMnO$_3$.
Then Section~\ref{sect:discussion} presents potential applications and extensions, and discusses limitations.
An advantage of the present Wannier-based perturbative scheme is that it is
built on already established workflows for computing phonons and magnetic
exchange parameters, and can therefore be integrated into existing
first-principles calculations. A standard ground-state DFT calculation first
provides the reference electronic structure. A DFPT phonon calculation then
yields the phonon frequencies and the variation of the self-consistent potential
$\partial V_{\text{SCF}}/\partial \tauu$, evaluated in the reference state.
Maximally localized Wannier functions are constructed to transform the Kohn--Sham Hamiltonian $H(\vec{k})$ into
its real-space representation $H(\vec{R})$~\cite{mostofi2008wannier90}. The EPC is then interpolated in the
Wannier basis to obtain $g_{ij,u}(\vec{R}_e,\rp)$~\cite{ponce2016epw}.
Finally, magnetic post-processing codes~\cite{TB2J} use the Wannier
Hamiltonian to build the magnetic Green's functions and evaluate the MFT
expressions, so that the four-point vertex can be integrated to obtain the SLCs
$\partial J/\partial \tauu$. The workflow is presented in
Fig.~\ref{fig:flow}.

\section{Formalism}
\label{sect:formalism}

To describe SLC, we first need (i) the electronic Hamiltonian of the reference state and (ii) the coupling between lattice displacements and the electron density. The remaining task is to combine these two ingredients to obtain the SLC parameters. The notation and conventions are summarized in Table~\ref{tab:notation}.

\begin{table}[t]
  \caption{Summary of notations used in the paper. Bold symbols are vectors, double bar symbols are matrices.}
  \label{tab:notation}
  \centering
  \begin{tabular}{@{}rl@{}}
    \toprule
    $n, m$ & electron band indices \\
    $i, j, k, l$ & electron Wannier orbital indices \\
    $I, J$ & atomic site indices (atom $I$ has orbitals $\{i\}$) \\
    $\vec{k}, \qs, \mathbf{q}_{\mathrm{P}}$ & wave-vectors (electron, magnon, and phonon) \\
    $\Ri$, $\Rj$ & cell vector of $I$, $J$, $\Rij=\Rj-\Ri$ \\
    $\mat{H}_{IJ}$ & tight-binding Hamiltonian (orbitals $i \in I, j \in J$) \\
    $\mat{U}_{jn,\vec{k}}$ & unitary WF transformation (orbital $j$, band $n$) \\
    $\epsilon$ & electron energy \\
    $\mat{G}_{IJ}(\epsilon)$ & Green's function (orbitals $i \in I, j \in J$). \\
    $\mat{p}_I, \mat{\Delta}_I$ & onsite exchange matrices at site $I$ (orbitals $i, i' \in I$) \\
    $g_{ij, u}(\re, \rp)$ & EPC (orbital $i, j$, displacement $u$) \\
    $u, v$ & displacement index (atom and Cartesian direction) \\
    $\alpha, \beta$ & 0 or directions ($ \alpha, \beta \in \{0, x, y, z\}$) \\
    $\msigma^\alpha$ & Pauli matrices ($\msigma^0$ is the $2\times 2$ identity matrix) \\
    \bottomrule
  \end{tabular}
\end{table}

\subsection{Green's function method for computing the Heisenberg parameters}
For the purely spin-dependent part of the Hamiltonian,  we need to compute the spin--spin interaction parameters. These parameters can be calculated with the MFT~\cite{LKAG} from local spin-density
approximation (LSDA)~\cite{von1972,vosko1980} within DFT. In the MFT method, the local spin rotation is treated as a perturbation, and
the variation of the band energy is calculated through a Green's function approach. The
method was originally used with local-basis methods~\cite{LKAG}.
In a DFT method with non-local basis set (e.g., plane-waves), this perturbation can be realized by
mapping the Kohn-Sham Hamiltonian to a localized WF basis set and
rotating the spin of the WFs~\cite{LKAGwannier}.

The electron Wannier function can be constructed using Bloch wave functions through:
 \begin{equation}
   \label{eq:wannbloch}
   \ket{j \vec{R}_j} = \sum_{n\vec{k}} e^{-i\vec{k}\cdot\vec{R}_j}\mat{U}_{jn,\vec{k}}\ket{n\vec{k}} \,,
 \end{equation}
where $\mat{U}_{jn}$ is a
 unitary transformation matrix that mixes the bands and localizes the electron orbitals~\cite{MLWFrmp,
   mostofi2008wannier90}. This allows us to express the electron Kohn-Sham Hamiltonian in the WF basis as $\mat{H}(\vec{R})$ (double-underlined symbols are matrices in WF orbital space).
 Consequently, the electron Green's function in $\vk$-space is $\mat{G}(\vec{k}, \epsilon)= \left[
   \epsilon \mat{I} -  \mat{H}(\vec{k})  \right]^{-1}$, where $\epsilon$ is the energy of the electron. In the following, we will drop the $\epsilon$ dependence for simplicity. The real-space Green's function is obtained by Fourier transform:
 $\mat{G}(\vec{R}) = \sum_{\vec{k}} e^{-i\vec{k} \cdot \vec{R}} \mat{G}(\vec{k})$.  We can decompose the full Green's function for each inter-site orbital pair $\gr_{ij}$ as the charge and three spin components,
 \begin{equation}
   \label{eq:T}
   \gr_{ij} =  \sum_{\alpha=0,x,y,z} G^{\alpha}_{ij} \msigma^{\alpha},
 \end{equation}
 where $\msigma^0$ is the $2\times 2$ identity matrix, and $\msigma^{x,y,z}$ are Pauli matrices.   Additionally, from the onsite block of $\mat{H}_{ij}$ at atom $I$, we extract the spin-splitting matrix $\mat{p}_I$ from the spin part of the Hamiltonian. This quantity acts as an effective exchange-correlation magnetic field in DFT, and the spin-rotation perturbation is realized by rotating it. Together with the Green's function, it allows us to compute the generalized exchange $\mat{A}_{IJ}$, a $4\times 4$ matrix defined as
 \begin{equation}
   \label{eq:defA}
   A_{IJ}^{\alpha\beta} =-\frac{1}{\pi} \int_{-\infty}^{E_F}   \Tr{\mat{p}_I\mat{G}_{IJ}^\alpha \mat{p}_J \mat{G}_{JI}^\beta}\ d\epsilon,
 \end{equation}
 where $\alpha, \beta \in  \{ 0, x, y, z\}$, $\Tr$ is the trace over the
 orbital space, and $\mat{p}_I$ is the matrix of $p_{ii'}$, where $i, i'$ are the orbital indices of atom $I$, and $\mat{G}_{IJ}$ is the Green's function matrix between the orbital indices of $i$ and $j$ with $i$ on atom $I$ and $j$ on atom $J$.  In a periodic system, we can choose the $IJ$ pair with $I$ in the
 cell at the origin, and $J$ in the cell translated by a lattice vector $\vec{R}_{J}$.

  \begin{equation}
    \label{eq:defA_periodic}
    A_{IJ}^{\alpha\beta}(\Rj ) =-\frac{1}{\pi} \int_{-\infty}^{E_F}   \Tr{\mat{p}_I\mat{G}_{IJ}^\alpha(\Rj) \mat{p}_J \mat{G}_{JI}^\beta(-\Rj)}\ d\epsilon,
  \end{equation}

From $\mat{A}_{IJ}$, we can compute the isotropic exchange $J^\mathrm{iso}$, the anisotropic exchange $\mat{J}^\mathrm{ani}$, and the DMI $\vec{D}$ using the following relations

 \begin{align}
   J^\mathrm{iso}_{IJ}&=\Im(A_{IJ}^{00}-A_{IJ}^{xx}-A_{IJ}^{yy}-A_{IJ}^{zz}) \label{eq:Jiso}, \\
   J_{IJ}^{\mathrm{ani},\alpha\beta} & = \Im(A_{IJ}^{\alpha\beta}+A_{IJ}^{\beta\alpha}) \label{eq:Jani}, \\
   D^{\alpha}_{IJ} &=  \Re (A_{IJ}^{0\alpha} - A_{IJ}^{{\alpha 0}})  \label{eq:DMI}.
 \end{align}

  In a collinear spin system without spin--orbit coupling (SOC), the DMI and anisotropic exchanges are
  zero, and the exchange parameter can be written in the LKAG formalism~\cite{LKAG, LKAGwannier} in the following way:

 \begin{equation}
   \label{eq:LKAGJ}
   J^\mathrm{iso}_{IJ}(\vec{R}_J)=\frac{1}{4\pi} \int_{-\infty}^{E_F}d\epsilon \Im \Tr{\mat{\Delta}_I\mat{G}_{IJ}^{\uparrow}(\vec{R}_J)\mat{\Delta}_J\mat{G}_{JI}^{\downarrow}(-\vec{R}_J)},
 \end{equation}
 where $\mat{\Delta}_{I} =
 \mat{H}_I^{\uparrow}(\vec{R}=0)-\mat{H}_I^{\downarrow}(\vec{R}=0)$.

 \subsection{Electron--phonon coupling}
In the case of the interaction between the electrons and the lattice vibrations, we need to obtain the EPC matrix elements. These matrix elements can be calculated through finite differences~\cite{monserrat2018electron} or density functional
  perturbation theory (DFPT)~\cite{gonzeDFPTpra, gonzeDFPTcg, gonzeDFPTparameters, phonondfpt, giustinoRMP}, by using an analytic form of
 the derivative of the potential~\cite{ScEPC1986, EPCLinres} and its wavevector dependence, which avoids the need for
 large supercells. Here we use the DFPT method,
 which is implemented in several DFT packages~\cite{abinit,giannozzi2009quantum}.

In this work, we make use of the Wannier representation of
EPC~\cite{PhysRevLett.98.047005, PhysRevB.76.165108} so that it can be combined
with the local-spin perturbation. The EPC parameter reads:
 \begin{equation}
   \label{eq:ephv}
     g_{mn, \nu} (\vec{k}, \vec{q}_\mathrm{P}) = \sqrt{\frac{\hbar}{2\omega_{\qp\nu}}}\bra{m \vec{k} + \vec{q}_\mathrm{P}} \dqpv V\ket{n\vec{k}},
 \end{equation}
  where $m$ and $n$ are electron band indices, $\nu$ is a phonon mode
  index, $\omega_{\qp\nu}$ is the phonon angular frequency, and $\vec{k}$ and $\vec{q}_\mathrm{P}$ are the electron and phonon wavevectors. The factor $\sqrt{\hbar/(2\omega_{\qp\nu})}$ ensures that $g_{mn,\nu}$ has units of energy.
 The EPC can be transformed into a Wannier
  representation~\cite{PhysRevB.76.165108}:
  \begin{equation}
    \label{eq:ephv_wannier}
    g_{ij, u} (\re, \rp) =\bra{i \vec{0}} \delta_{\tauu} V\ket{j\re}
 \end{equation}
 where $i$ and $j$ are the indices of electron Wannier
functions, and $u$ is the displacement index. The vectors $\re$ and $\rp$ locate
the electron and displaced atom, respectively.


 The transformation from the Bloch to Wannier space
 representation  can be done with:
 \begin{equation}
   \label{eq:epcwannrr_full}
   \begin{aligned}
   g_{ij,u}(\re, \rp) = &\frac{1}{N_{\qp} N_\vk}  \sum_{\vk, \qp} e^{-i(\vk \cdot \re + \qp \cdot \rp)} \\
    &\times \sum_{mn\nu} U^{*}_{im,\vk+\qp}\, g_{mn,\nu}(\vk, \qp)\, U_{nj,\vk}\, u^{-1}_{\nu u,\qp}
   \end{aligned}
 \end{equation}
  where $N_{\qp}$ is the number of $\vec{q}_\mathrm{P}$ points, $N_\vk$ is the number of $\vec{k}$ points, the $U$ matrices are  the
 unitary matrices which rotate the electron Bloch states into the Wannier
 functions, and $u^{-1}_{\nu u,\qp}$ projects phonon eigenmode $\nu$ onto Cartesian displacement $u$.

When we consider the collinear magnetic case, the treatment is analogous but now we have spin-resolved EPC $g^\uparrow_{ij,u}(\vec{R}_\mathrm{e},\vec{R}_\mathrm{p})$ and $g^\downarrow_{ij,u}(\vec{R}_\mathrm{e},\vec{R}_\mathrm{p})$ constructed with Wannier functions for each spin manifold, and the perturbation potentials $\delta_{\tauu}V^\uparrow$ and $\delta_{\tauu}V^\downarrow$, respectively.

 \subsection{Mixing the spin and phonon perturbation}
\label{sect:mixing_SLC}

 \begin{figure}[htbp]
   \centering
   \includegraphics[width=\linewidth]{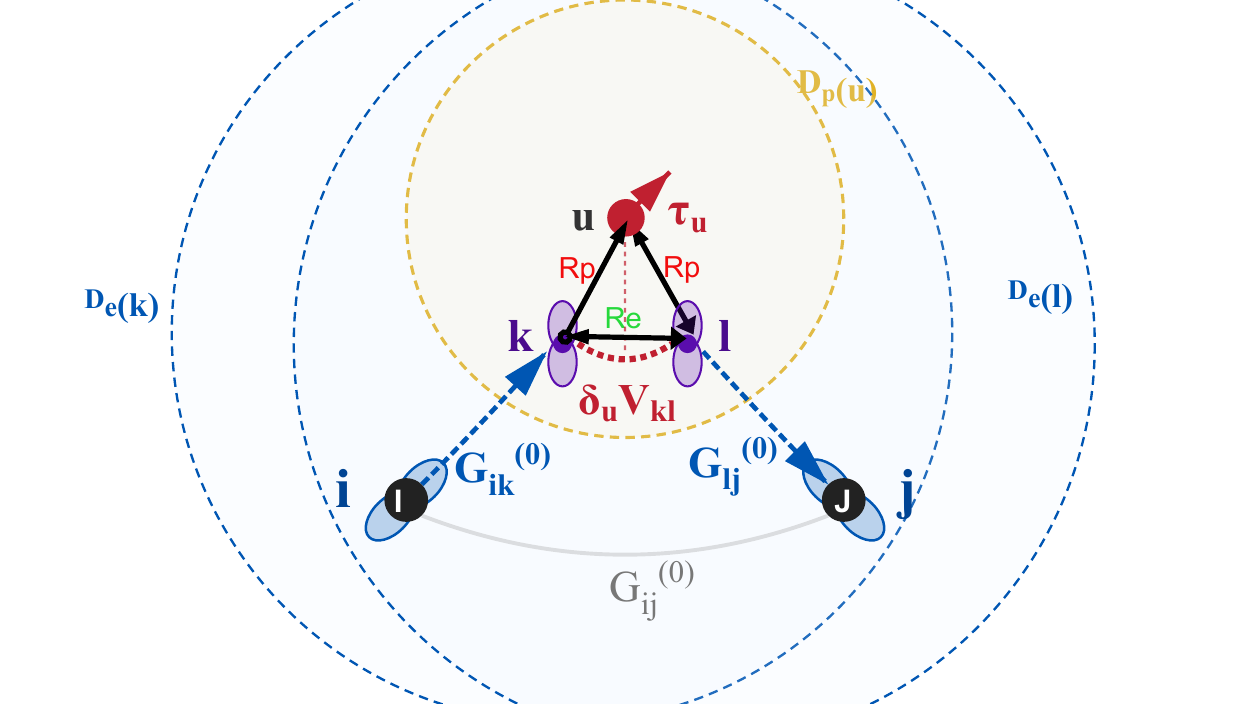}
    \caption{Schematic diagram of electron scattering due to a lattice
      distortion. An electron propagates from Wannier function $i$ (centered at atom $I$) to $j$ (centered at atom $J$) via intermediate states $k$ and $l$. The scattering is caused by the variation in the potential, $\delta_{\tauu} V_{lk}$, induced by the displacement $\tauu$ of atom $u$. The blue arrows represent the unperturbed Green's functions $G_{ik}^{(0)}$ and $G_{lj}^{(0)}$. The dashed circles illustrate the spatial locality of the interactions: $D_\mathrm{e}$ indicates the electron interaction range (cutoff distance for electron pairs), and $D_\mathrm{P}$ indicates the electron--lattice interaction range centered at the displaced atom. $D_{\mathrm{e}}(k)$, $D_{\mathrm{e}}(l)$, $D_{\mathrm{P}}(u)$ are the interaction range around $k$, $l$, and $u$, respectively.}
   \label{fig:epgreendiagram}
 \end{figure}
Once we have access to both the exchange parameters and the EPC, we can construct the SLC. In this work, we apply an idea similar to the MFT~\cite{LKAG} approach but for a lattice distortion instead of a spin rotation. Thus, we compute the variation of the exchange parameter $\mat{A}_{IJ}$ by considering the change of the electronic Green's function under a lattice distortion to obtain the SLC parameters.

 Formally, this means that the electron Green's
 function $G_{ij}(\rj)$ is perturbed by a displacement $\tauu$ of atom $u$ in the cell $\vec{R}_{u}$, which changes the hopping between all orbital pairs
 $\vec{R}_k$ and $\vec{R}_l$, as shown in Fig.~\ref{fig:epgreendiagram}, in which the cell vectors are not shown. Hence, the first-order perturbed Green's function can be written as:
 \begin{multline}
   \label{eq:greenpert}
   \delta_{\tauu}G_{ij}(\rj)= \sum_{k, l} G^{(0)}_{ik}(\ri-\mathbf{R}_k) \\ \times g_{kl,u}(\mathbf{R}_l-\mathbf{R}_k, \vec{R}_{u}-\mathbf{R}_k) G^{(0)}_{lj}(\rj-\mathbf{R}_l) \,,
\end{multline}
where $G^{(0)}$ is the Green's function without the phonon perturbation. 
This allows us to compute the variation of $\mat{A}_{IJ}$ in Eq.~\eqref{eq:defA} with respect to the atomic displacement amplitude, which can be written as the sum of four parts:
\begin{equation}
  \label{eq:dAdvqp}
\begin{aligned}
 \delta_{\tauu} A_{IJ}^{\alpha\beta}(\Rj)  = & -\frac{1}{\pi} \Tr \int_{-\infty}^{E_{F}} d\epsilon \\
                                                  &  \phantom{+} \mat{p}_I \delta \mat{G}_{IJ}^{\alpha} (\Rj) \mat{p}_J \mat{G}_{JI}^{(0),\beta}(-\Rj) \\
                                                  & + \mat{p}_I \mat{G}_{IJ}^{{(0),\alpha}}(\Rj)\mat{p}_J  \delta \mat{G}_{JI}^{\beta} (-\Rj)\\
                                                  & + \delta\mat{p}_{I}  \mat{G}_{IJ}^{(0),\alpha}(\Rj) \mat{p}_J \mat{G}_{JI}^{(0),\beta}(-\Rj) \\
                                                  & + \mat{p}_I\mat{G}_{IJ}^{(0),\alpha}(\Rj)\delta \mat{p}_J \mat{G}_{JI}^{(0),\beta}(-\Rj),
\end{aligned}
\end{equation}
where $\delta\mat{p}_I$ is the onsite spin-splitting amplitude change due to the
atomic displacement $\tauu$, and $\mat{G}^{(0), \alpha}$ is the Pauli component $\alpha$ of the unperturbed Green's function matrix between sites $I$ and $J$ [Eq.~\eqref{eq:T}]. The trace $\mathrm{Tr}$ is taken over the orbitals of one atom. The first two terms describe the lattice-induced change of the two Green's-function propagators, while the last two terms describe the displacement-induced change of the onsite spin splittings.
Note that we omit the subscript $\tauu$ in Eq.~\eqref{eq:dAdvqp} for concision.
By inserting Eq.~\eqref{eq:greenpert} into Eq.~\eqref{eq:dAdvqp}, the latter can be
evaluated. The derivatives of $J^\mathrm{iso}$, $\Jani$ and $\vec{D}$ can thus be obtained with a
similar mapping of Eqs.~\eqref{eq:Jiso}--\eqref{eq:DMI}. In this work, we focus on the collinear-spin case. 

In the collinear spin case, without SOC, the exchange parameter becomes:
\begin{equation}
  \label{eq:dJdvqp}
\begin{aligned}
 \delta_{\tauu} J_{IJ}^\mathrm{iso}(\Rj)  = & \frac{1}{4\pi}\Im\Tr\int_{-\infty}^{E_{F}} d\epsilon \\
                                                  &  \phantom{+} \mat{\Delta}_I \delta \mat{G}_{IJ}^{\uparrow} (\Rj) \mat{\Delta}_J \mat{G}_{JI}^{(0),\downarrow}(-\vec{R}_J) \\
                                                  & + \mat{\Delta}_I \mat{G}_{IJ}^{{(0),\uparrow}}(\Rj)\mat{\Delta}_J  \delta \mat{G}_{JI}^{\downarrow} (-\Rj)\\
                                                  & + \delta\mat{\Delta}_I  \mat{G}_{IJ}^{(0),\uparrow}(\Rj) \mat{\Delta}_J \mat{G}_{JI}^{(0),\downarrow}(-\Rj) \\
                                                  & + \mat{\Delta}_I\mat{G}_{IJ}^{(0),\uparrow}(\Rj)\delta \mat{\Delta}_J \mat{G}_{JI}^{(0),\downarrow}(-\Rj), 
\end{aligned}
\end{equation}
with $\delta \Delta_I$ being the change in the spin splitting due to the phonon perturbation. We note that at the collinear level, the potential $V$ acting on the two spin channels is different and, as a result, the change in exchange splitting does not need to be zero.

 The advantage of this approach is that we can exploit the shortsightedness of the
electron hopping and electron--phonon interaction to limit the range of
$\vec{R}_l$ and $\vec{R}_k$ in Eq.~\eqref{eq:greenpert}. In practice, the range of the EPC is defined by the coarse $\vec{k}$ and $\qp$ grid, which gives the number of electron and lattice Wigner-Seitz cells. With sufficiently dense $\vec{k}$ and $\qp$ grids, the EPC should decay and vanish outside the Wigner-Seitz supercell, allowing the range of the SLC parameters to be converged systematically.

Finally, we note that in this framework, the SLC parameters describe the effective interaction between localized spins mediated by lattice distortions of the electronic system. Thus, extracting SLC from EPC can be viewed as a downfolding (DF) procedure.
In practical calculations, the required inputs are the Wannier Hamiltonian, the
onsite spin-splitting matrices, the unperturbed Green's functions, and the
Wannier-represented EPC matrix. From these quantities, the method directly
returns $dJ_{IJ}/d\tauu$ for selected magnetic pairs $IJ$ and atomic
displacements $\tauu$, and the same expressions can be resolved into
orbital-pair contributions.



\section{Application to $\mathrm{SrMnO}_3$}
\label{sect:example}

\begin{figure}[htbp]
  \centering
  \includegraphics[width=0.4\textwidth]{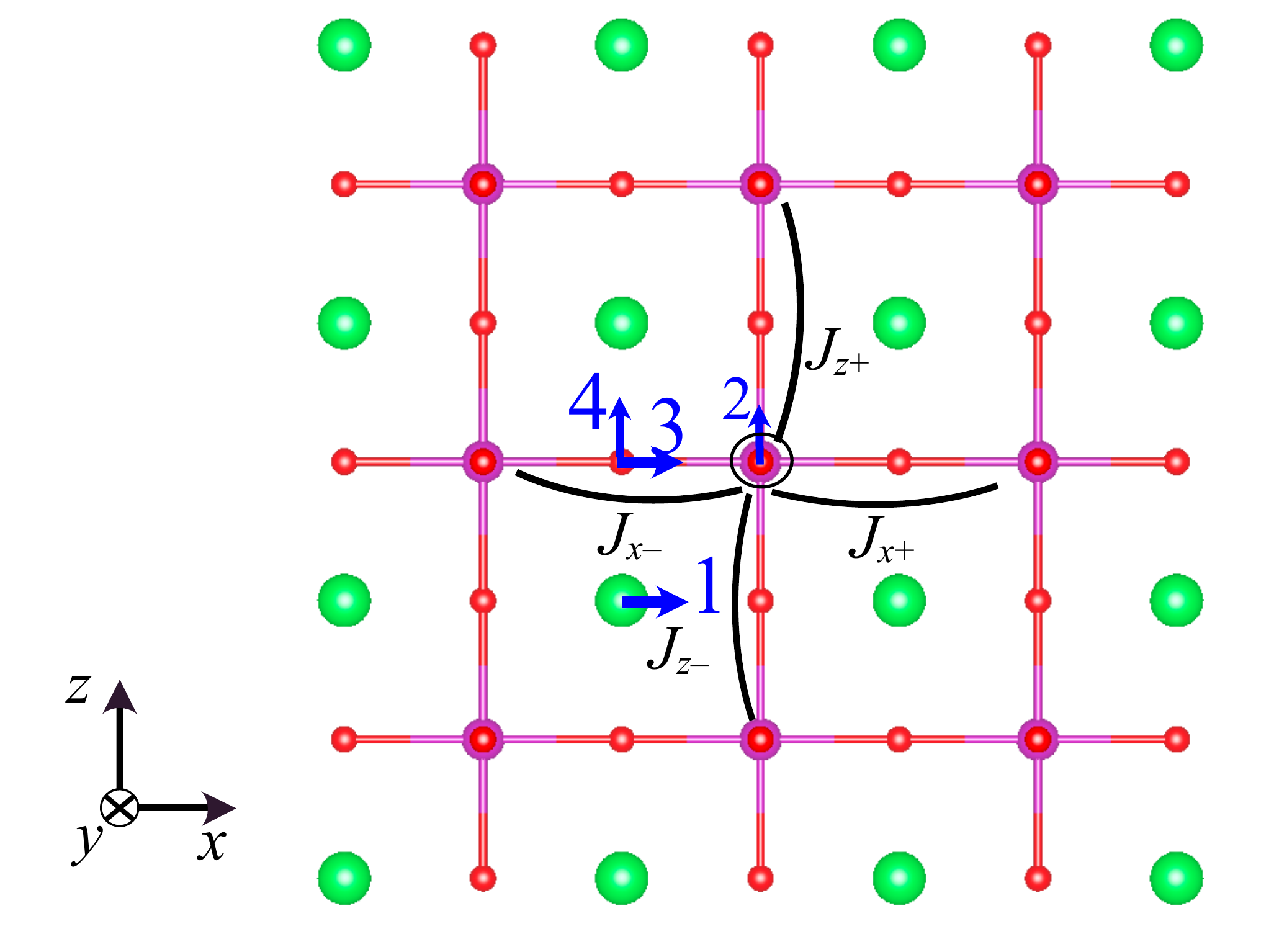}
  \caption{Structure of SrMnO$_3$. The green, purple, and red dots are
    the Sr, Mn, and O atoms, respectively. The four atomic displacements labeled
    1--4 correspond to Sr, Mn and O1, O2 displacements along the crystallographic directions, as defined in the text. The exchange couplings with the
    central Mn atom (inside the black circle) are discussed in the text, e.g.,
   $J_{x+}$ is the exchange parameter between the Mn and its first nearest neighbor in the positive $x$ direction. }
  \label{fig:SMO}
\end{figure}

We validate the method with SrMnO$_3$ (SMO) in the cubic perovskite structure, where the magnetic exchange between
Mn $3d$ electrons is strongly coupled to the phonons
\cite{SMOstrain,SMOspinphon}. The Mn
ions are in a $3d^3$ electronic configuration, and interact with their first Mn neighbors through a
superexchange mechanism, transmitted through the Mn-O orbital
hybridization, and modulated by the Mn-O-Mn bonds.

The DFT calculations were performed with the \texttt{Quantum ESPRESSO} package~\cite{giannozzi2009quantum}. We make use of the
PBEsol Generalized Gradient Approximation (GGA)~\cite{PBEsol} exchange-correlation functional, and
norm-conserving pseudopotentials from \texttt{PseudoDojo}~\cite{dojo,ONCVPSP}. The phonons and
EPC matrices are calculated with DFPT, with a $5\times
5\times 5$ Monkhorst-Pack grid of $\qp$-points, and a $10\times 10 \times 10$ grid of
$\bf k$-points. The SMO cubic structure in a
ferromagnetic configuration is taken as the reference state.
The \texttt{Wannier90}~\cite{mostofi2008wannier90} package allows us to build maximally localized Wannier functions~\cite{MLWFrmp}, which include the Mn $3d$ and O $2p$ states.
We then carry out a DFPT calculation to compute the phonons and
EPC. Based on these results, the Wannier representation of the
EPC is constructed with the \texttt{EPW} code~\cite{carrascoalvarez2025,ponce2016epw, PhysRevB.76.165108}. In the current implementation of \texttt{EPW}, the
Wannier representation of the EPC can be calculated at the collinear level, obtaining a spin-resolved EPC $g^\uparrow$ or $g^\downarrow$~\cite{carrascoalvarez2025}. We did not include the Hubbard $U$ correction in this work for simplicity, and it is left for future work since it has recently been made available in \texttt{EPW}~\cite{Yang2025,CarrascoAlvarez2026}.
For the calculation of the exchange coupling parameters, we use the magnetic
force theorem~\cite{LKAG} implemented in the \texttt{TB2J}~\cite{TB2J} code based on the Wannier
function Hamiltonian.

\begin{figure}
    \centering
    \includegraphics[width=\linewidth]{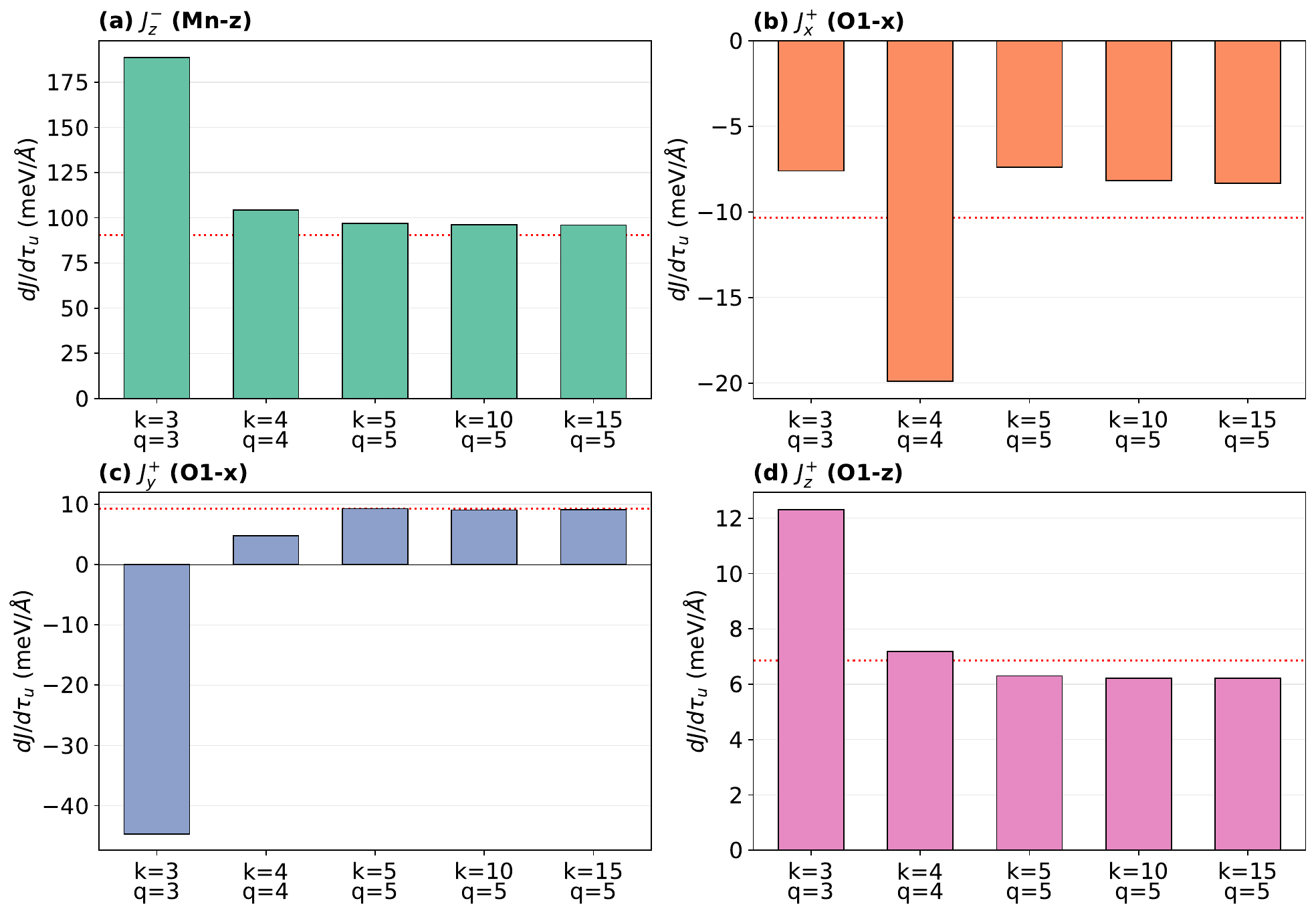}
    \caption{Convergence of selected SLC parameters with respect to the
      $\mathbf{k}$- and $\qp$-point meshes used in the electron--phonon DFPT calculation,
      denoted by $n_\vk \times n_\vk \times n_\vk$ and
      $n_{\qp} \times n_{\qp} \times n_{\qp}$, respectively. The red dotted lines show the
      values obtained from the finite-difference $3\times 3 \times 3$
      supercell calculation, which is used as a consistency check rather than as
      an exact reference.}
    \label{fig:dJdxconv}
\end{figure}

To benchmark the method, we use a finite-difference method (FD) to compute the
SLC parameters for comparison. We displace the atoms by $\pm 0.02$
\AA{} along the $x$, $y$, and $z$ directions, and compute the exchange
parameters in a $3\times 3 \times 3$ supercell. Due to finite displacement effects, we find that the corresponding 3$\times$3$\times$3 $\qp$-point grid in PT is not sufficient and at least a 4$\times$4$\times$4  $\qp$-point grid is needed, see Fig.~\ref{fig:dJdxconv}. The PT approach also offers a significant computational advantage over the FD
method. 

For SrMnO$_3$, each FD calculation in the $3\times 3 \times 3$ supercell requires about 230~CPU-hours (AMD EPYC Milan 2.45 GHz), and two calculations (positive and negative displacements) are needed for each perturbation. This yields a total FD cost of $t_{\rm FD} \approx 460\times n$~CPU-hours, where $n$ is the number of independent perturbations. In contrast, the DFPT calculation with a $10\times 10 \times 10$ $\vec{k}$-point mesh and a $5\times 5 \times 5$ $\qp$-point mesh requires about 800~CPU-hours as a one-time cost. 

The PT calculation corresponds to an effective $5\times 5 \times 5$ supercell (from the $\qp$-point grid), which would be even more CPU-time and memory demanding if done with the FD approach. 

A more important advantage of the PT calculation is the memory usage.
Because the DFPT calculation is performed independently for each $\qp$ point,
the memory footprint does not grow directly with the number of $\qp$ points,
apart from small post-processing overheads. Some $\qp$ points reduce the number
of symmetries and can therefore increase the maximum memory usage among the
individual calculations. In contrast, for the supercell calculation, the storage
of the wave functions scales as $O(N^2)$, where $N$ is the supercell volume. In
the present case, the $3\times3\times3$ calculation requires about 40 GB of
memory, while a $5\times 5\times 5$ supercell would require about 21 times more
memory, or roughly 880 GB.

\begin{figure}[htbp]
  \centering
  \includegraphics[width=0.45\textwidth]{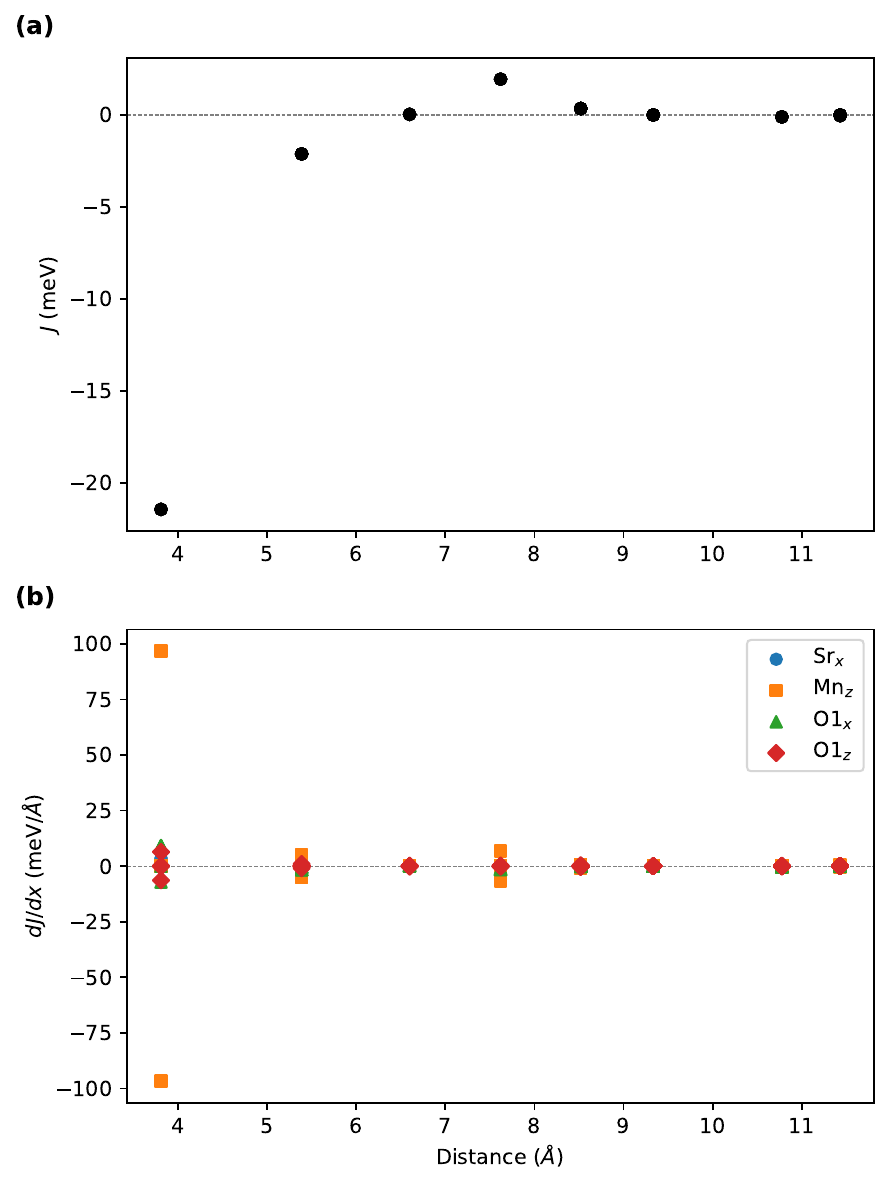}
  \caption{(a) Exchange parameters and (b) SLC parameters as functions of
    Mn--Mn distance for the atomic displacements shown in Fig.~\ref{fig:SMO}.
    The sign convention is that negative $J$ corresponds to antiferromagnetic
    exchange.}
  \label{fig:dJvsR}
\end{figure}

\begin{figure}[htbp]
  \centering
  \includegraphics[width=0.45\textwidth]{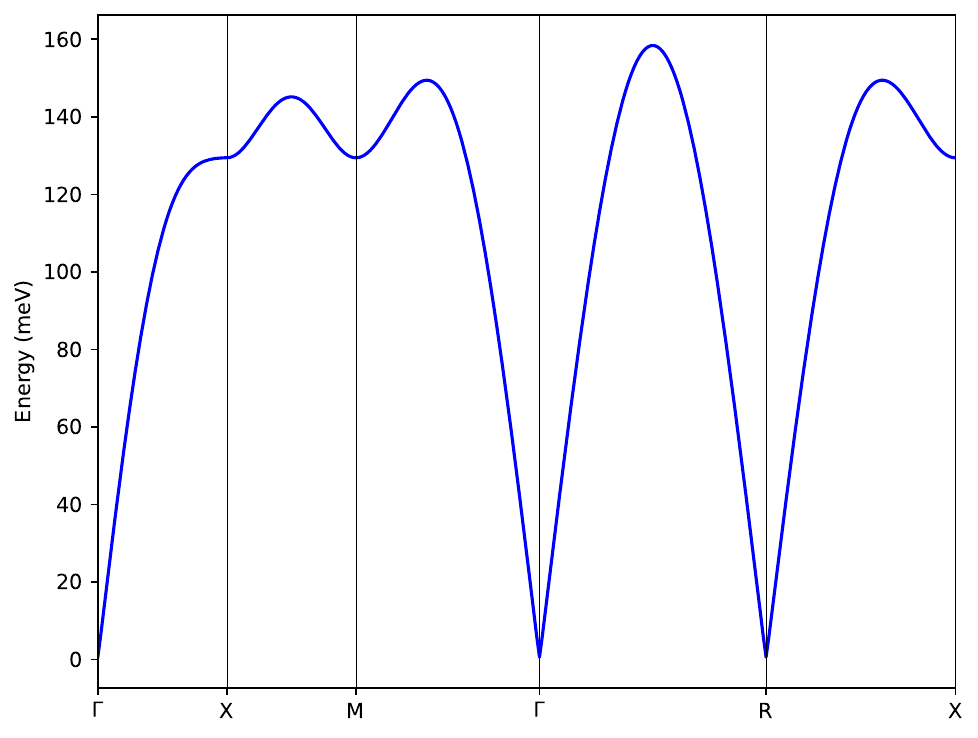}
  \caption{Magnon band structure of SrMnO$_3$ computed from the exchange parameters in the Brillouin zone (BZ) of the cubic structure. The reference state is the G-type antiferromagnetic (AFM) ground state, in which each Mn moment is antiparallel to all nearest-neighbor Mn moments, which doubles the unit cell by a supercell matrix of $\mat{M}_{SC}$ of ((0,$\frac{1}{2}$,$\frac{1}{2}$),($\frac{1}{2}$,0,$\frac{1}{2}$),($\frac{1}{2}$,$\frac{1}{2}$,0)).  The $q_{\mathrm{P}}$ at $R$ ($\frac{1}{2}$, $\frac{1}{2}$,$\frac{1}{2}$) has zero frequency, as the wave-vector is folded to $\vec{q}_{\mathrm{P}}\cdot \mat{M}_{SC}=(1,1,1)$ of the supercell BZ, equivalent to the $\Gamma$ point. }
  \label{fig:magnon_bands}
\end{figure}

We validate the computation of the SLC parameters with four
atomic distortions: one Sr displacement, one Mn displacement, and two O atom displacements (along and perpendicular to the
Mn-O bond in the $a$ direction as shown in Fig.~\ref{fig:SMO}), labeled by Sr-$x$, Mn-$z$, O-$x$, and O-$z$, respectively. The exchange and the
SLC parameters connecting the Mn atoms are plotted
as a function of the distance.
The first-neighbor $J$ is significantly larger than the longer-ranged ones. The
derivatives of $J$ with respect to atomic displacements also decay with distance as shown in Fig.~\ref{fig:dJvsR}. These exchange parameters can be used to compute the magnon band structure via linear spin-wave theory (LSWT)~\cite{toth2015linear} as implemented in \texttt{TB2J}. The resulting magnon dispersion, shown in Fig.~\ref{fig:magnon_bands}, shows a typical antiferromagnetic (AFM) behavior which is linear close to the zone center. We note that although our DFT calculations are done within a unit cell of FM structure, the computation of the magnon band for the reference AFM ground-state structure can be done following the method in Ref.~\cite{toth2015linear}.

The exchange values in Fig.~\ref{fig:dJvsR} are much larger than those reported in previous works~\cite{zhu2020magnetic}. This difference is mainly due to the absence of a Hubbard $U$ correction in the calculations used for the present demonstration. For comparison, a DFT+$U$ calculation with $U(\mathrm{Mn})=3$~eV and $J(\mathrm{Mn})=0$~eV gives a first-neighbor Mn--Mn exchange of about $-7.04$~meV, whereas the calculation without $U$ used here gives $-21.45$~meV. The Hubbard correction pushes the occupied and unoccupied Mn $d$ states further apart, reducing Mn $d$--O $p$ hybridization and thus the superexchange interaction. The SLC parameters are expected to be overestimated for the same reason. As we mentioned before, the inclusion of the $U$ is left for future studies and we mainly focus on demonstrating the method to compute the SLC parameters regardless of the value of $J$. The corresponding values of the derivative of the isotropic exchange $J^\mathrm{iso}$ with respect to the displacements are shown in Table~\ref{tab:dJ}.
\begin{table}[b]
  \centering
  \begin{tabular}{c|cccccc}
    \toprule
   {distortion} & $J_{x+}$ & $J_{x-}$ & $J_{y+}$ & $J_{y-}$ & $J_{z+}$& $ J_{z-}$  \\
    $dJ/d\tauu$ (meV/\AA) & \multicolumn{6}{c}{} \\
    Sr-$x$ PT  & -0.06 & 0.01 & -0.56 & 3.41 & -0.56 & 3.41 \\
    Sr-$x$ FD  & -0.44 & -0.05 & -0.41 & 3.27 & -0.41 & 3.27 \\

    Mn-$z$ PT & -0.01 & -0.01 & -0.01 & -0.01 & -96.57 & 96.87 \\
    Mn-$z$ FD  & -0.01 & -0.01 & -0.01 & -0.01 & -90.51 & 90.46 \\

    O1-$x$ PT  & -7.38 & -0.10 & 9.20 & 9.20 & 9.20 & 9.20 \\
    O1-$x$ FD  & -10.34 & -0.13  & 9.31 & 9.31 & 9.31 & 9.31 \\

    O1-$z$ PT  & -0.00 & -0.02 & 0.00 & 0.00 & 6.48 & -6.48 \\
    O1-$z$ FD  & 0.05 & 0.05 & -0.05 & 0.08 & 6.86 & -6.84 \\
  \end{tabular}
  \caption{First-neighbor SLC parameters $dJ/d\tauu$ for the displacements
    labeled in Fig.~\ref{fig:SMO}. PT denotes the perturbative method developed
    here, and FD denotes the finite-difference supercell calculation used as a
    consistency check. The exchange sign convention is
    $H^\mathrm{spin}=-\sum_{IJ}J_{IJ}\Si\cdot\Sj$, so negative $J$ corresponds
    to antiferromagnetic exchange. The units are meV/\AA.\label{tab:dJ}}
\end{table}
\begin{figure}
  \centering
  \includegraphics[width=\linewidth]{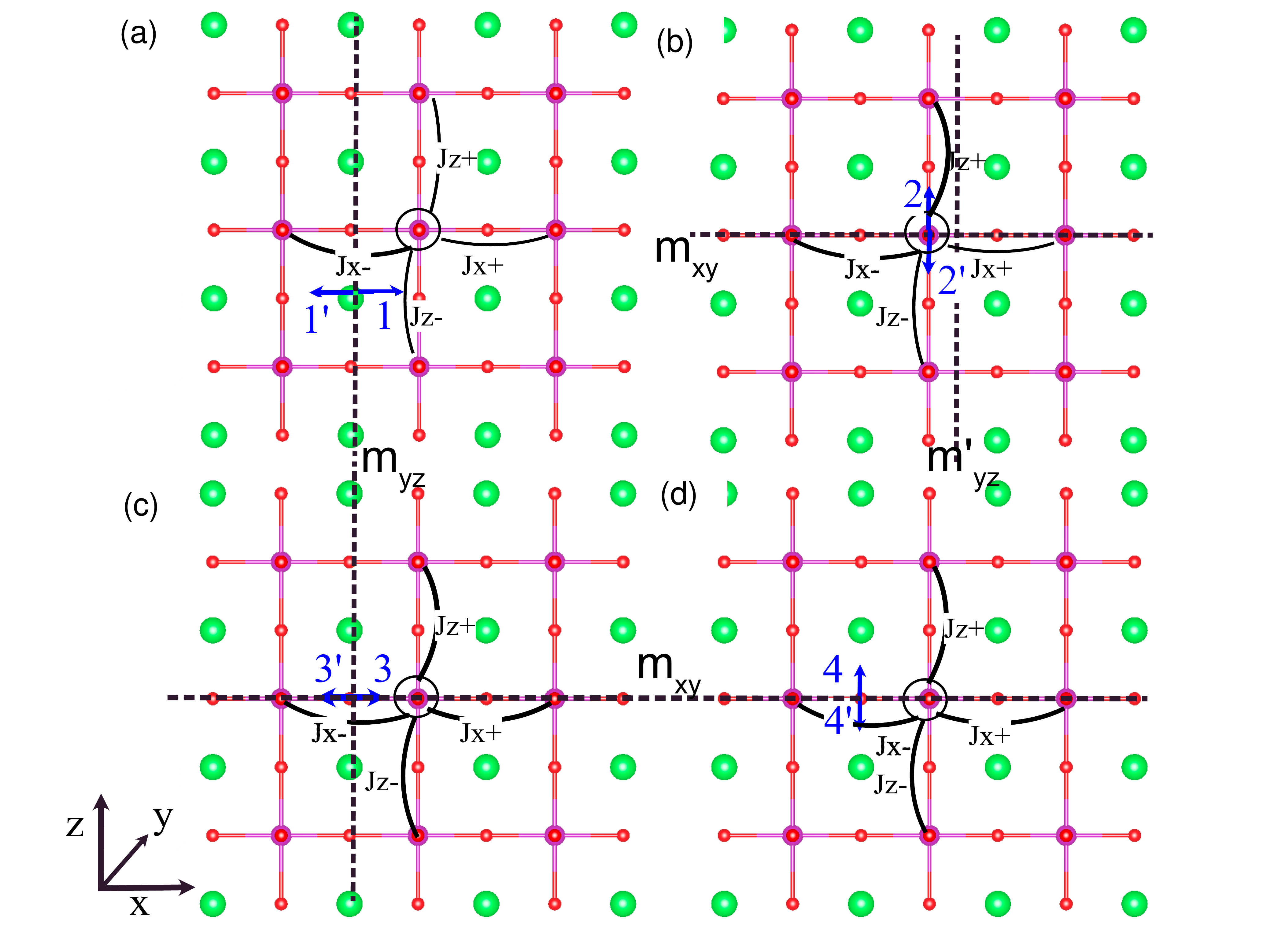}
  \caption{Symmetry relations for the SLC parameters in SrMnO$_3$: (a) Sr-$x$ displacement, (b) Mn-$z$ displacement, (c) O1-$x$ displacement, and (d) O1-$z$ displacement. These panels correspond to displacements 1--4 in Fig.~\ref{fig:SMO}. The dashed lines indicate the mirror planes. The Sr atoms are not in the same plane as the Mn and O atoms.}
  \label{fig:SMOsymmetry}
\end{figure}

For the Sr-$x$ distortion [Fig.~\ref{fig:SMOsymmetry}(a)], displacements 1 and $1'$ are related by the mirror plane $m_{yz}$. They therefore give the same change in $J_{x-}$, so the first derivative of $J_{x-}$ with respect to the Sr-$x$ displacement vanishes. The couplings $J_{y+}$ and $J_{y-}$ are not symmetry equivalent for this distortion, whereas $J_{y-}$ and $J_{z-}$ are related by a four-fold rotation around the $x$ axis through the displaced Sr atom (not shown), so their derivatives are equal.

For the Mn-$z$ distortion [Fig.~\ref{fig:SMOsymmetry}(b)], displacements 2 and $2'$ are related by the mirror plane $m_{xy}$ and change the in-plane exchange couplings in the same way. As a result, the derivatives of $J_{x+}$, $J_{x-}$, $J_{y+}$, and $J_{y-}$ vanish, while the derivatives of $J_{z+}$ and $J_{z-}$ are antisymmetric.
For the distortion O-$x$ [Fig.~\ref{fig:SMOsymmetry}(c)], the first derivative of $J_{x-}$ should be 0, as the two displacements 3 and $3'$ are related by a mirror plane $m_{yz}$ and thus give the same effect on $J_{x-}$.
 The other SLCs for the 1NN are
non-zero. The $J_{y-}$, $J_{y+}$, $J_{z-}$, $J_{z+}$ should be equivalent by
a four-fold rotation symmetry, and indeed we obtain the same values.

For the distortion O-$z$ [Fig.~\ref{fig:SMOsymmetry}(d)], the O atom is displaced along the $z+$ direction. The $J_{x+}$ and $J_{x-}$ should be 0, as the two displacements 4 and $4'$ are related by a mirror plane $m_{xy}$ and thus give the same effect on $J_{x-}$. The same holds for $J_{y+}$ and $J_{y-}$. The $J_{z+}$ and $J_{z-}$ are anti-symmetric.

The results in Table~\ref{tab:dJ} confirm all the symmetry relations discussed above, validating the method, except for small numerical noise. We note that the results from the FD method have larger numerical noise when checking the symmetry relations. This is likely due to inaccuracies in the Wannier functions constructed for the distorted structures, which can break the symmetry.


As an additional numerical check, we calculate the exchange parameters in supercells with atomic
As an additional numerical check, we calculate the exchange parameters in supercells with atomic displacements and then obtain the derivatives of the $J$ values by FD. The qualitative agreement between PT and FD is good, but some quantitative differences remain (e.g., for O1-$x$, PT gives $-7.38$~meV/\AA{} while FD gives $-10.34$~meV/\AA). The discrepancies can be attributed to the following reasons: (i) the Wannier functions do not preserve their shape under finite displacements, (ii) numerical noise in the Wannier Hamiltonian, and (iii) wrapping of the contributions from periodic images in the FD supercell calculation. Regarding point~(3), the FD calculation was done in a $3\times 3 \times 3$ supercell due to memory limitations, while the PT calculation corresponds to an effective $5\times 5 \times 5$ supercell (from the $\qp$-point grid). The periodic images in the smaller FD supercell introduce spurious interactions that affect the SLC values. This suggests that the PT method is not only faster, but can also be more accurate than the FD approach, as it effectively bypasses the finite-size supercell errors by working directly in the primitive cell with a dense $\qp$-point grid.

%
%
%

Equations~\eqref{eq:defA} and~\eqref{eq:dAdvqp} can be evaluated for each pair of
orbitals rather than only for each pair of atoms. This decomposes both the total
exchange parameter and the SLC into orbital-pair contributions, providing a
direct way to identify the electronic channels responsible for the coupling.
We illustrate this orbital resolution for the Mn-$z$ displacement. For each Mn
atom, the active $3d$ orbitals are $d_{z^2}$, $d_{xz}$, $d_{yz}$,
$d_{x^2-y^2}$, and $d_{xy}$. The orbital-resolved isotropic exchange for
$J_{z+}=-21.45$~meV can therefore be written as the following $5\times 5$
matrix:
\begin{equation*}
  \begin{pmatrix}
    0.85 & 0.00   & 0.00   & 0.00  & 0.00  \\
    0.00 & -11.10 & 0.00   & 0.00  & 0.00  \\
    0.00 & 0.00   & -11.10 & 0.00  & 0.00  \\
    0.00 & 0.00   & 0.00   & -0.03 & 0.00  \\
    0.00 & 0.00   & 0.00   & 0.00  & -0.06
  \end{pmatrix}
\end{equation*}
The exchange is mainly due to the $d_{z^2}-d_{z^2}$, $d_{xz}-d_{xz}$, and
$d_{yz}-d_{yz}$ pairs. The latter two show a superexchange picture, as the
$d_{xz}$ and $d_{yz}$ orbitals are singly occupied and hybridize with the bridging oxygen
$2p$ orbitals. As the unoccupied $d_{z^2}$
orbital hybridizes with the oxygen $p_z$ orbitals, the hopping favors a
ferromagnetic alignment. Thus the antiferromagnetic interaction is the net result of
these two competing mechanisms. The contributions from other orbital pairs are
either zero due to symmetry or much smaller.

With the Mn-$z$ displacement, the distance between the two Mn atoms decreases. We
obtain the following derivative of $J_{z+}$:
\begin{equation*}
  \begin{pmatrix}
     6.21 & 0.00   & 0.00   & 0.00 & 0.00  \\
     0.00 & -51.37 & 0.00   & 0.00 & 0.00  \\
     0.00 & 0.00   & -51.37 & 0.00 & 0.00  \\
     0.00 & 0.00   & 0.00   & 0.07 & 0.00  \\
     0.00 & 0.00   & 0.00   & 0.00 & -0.11
  \end{pmatrix}
\end{equation*}
Both the ferromagnetic interaction between the $d_{z^2}-d_{z^2}$ pair, and the
antiferromagnetic interaction from the
  $d_{xz}-d_{xz}$, and
$d_{yz}-d_{yz}$ pairs are enhanced as the Mn-O-Mn bond is shortened. In total,
$J_{z+}$ becomes more antiferromagnetic. The contributions from other
orbital pairs are much smaller, or 0 due to the symmetry.

Another example is the $J$ between a  Mn-Mn pair along
the $y+$ direction  with the O1-$x$ displacement (displacement 3 in Fig.~\ref{fig:SMO} ). The isotropic exchange $J_{y+}$ is
$J^\mathrm{iso} = -21.45$ meV at a distance of 3.810 \AA. The orbital-resolved
contribution to $J_{y+}$ is:
\begin{equation*}
  \begin{pmatrix}
    0.20  & 0.00  & 0.00   & 0.00 & 0.00   \\
    0.00  & -0.06 & 0.00   & 0.00 & 0.00   \\
    0.00  & 0.00  & -11.10 & 0.00 & 0.00   \\
    0.00  & 0.00  & 0.00   & 0.64 & 0.00   \\
    0.00  & 0.00  & 0.00   & 0.00 & -11.10
  \end{pmatrix}
\end{equation*}
The derivative with respect to O1-$x$ displacement is $dJ_{y+}/d\tauu = 9.20$
meV/\AA, with orbital contributions:
\begin{equation*}
  \begin{pmatrix}
    0.03  & 0.00  & 0.00  & 0.25  & 0.00  \\
    0.00  & -0.01 & 0.00  & 0.00  & 0.00  \\
    0.00  & 0.00  & -2.51 & 0.00  & 0.00  \\
    -0.21 & 0.00  & 0.00  & -1.44 & 0.00  \\
    0.00  & 0.00  & 0.00  & 0.00  & 13.09
  \end{pmatrix}
\end{equation*}

Interestingly, the $J_{y+}$ has significant contributions from the
$d_{xy}-d_{xy}$ and $d_{yz}-d_{yz}$ pairs, both contributing -11.10 meV to
$J_{y+}$. However, their derivatives with respect to the O1-$x$ displacement are
very different: the $d_{yz}-d_{yz}$ pair enhances the antiferromagnetic
interaction by -2.51 meV/\AA, while the $d_{xy}-d_{xy}$ pair reduces the
antiferromagnetic interaction by 13.09 meV/\AA.

This can be understood as follows: with the O1 atom moving along the $x+$
direction, it gains stronger hybridization with the Mn atom, especially with the
Mn $d_{xy}$ orbital. This competes with the hopping along the Mn $d_{xy}$--O
$p$--Mn$_2$ $d_{xy}$ superexchange path in the $y$ direction, thus reducing the
antiferromagnetic interaction. On the other hand, the O1 displacement does not
significantly change the hybridization along the Mn $d_{yz}$--O $p$--Mn
$d_{yz}$ hopping path. There are also contributions from other orbital pairs,
notably the symmetric contribution from the $d_{x^2-y^2}-d_{z^2}$ and
$d_{z^2}-d_{x^2-y^2}$ pairs, which are affected by the O1 displacement but
largely cancel out (see the off-diagonal terms in the derivative matrix).

The analysis above shows that the orbital-resolved SLC parameters can provide deep insight into the mechanism of the SLC, and clearly illustrates the impact of ligand atoms in modulating the superexchange interaction.

\section{Discussion}
\label{sect:discussion}

This new method allows SLC parameters to be
computed by post-processing standard DFT and DFPT calculations. Its inputs are
the EPC matrix, the WF $U_{jn,\vk}$ matrix, and the Wannier Hamiltonian. These
quantities are already generated in established workflows for phonons,
electron--phonon coupling, and magnetic exchange parameters. The present approach
therefore adds the SLC calculation as a post-processing layer on top of these
well-established methods and software packages.

The algorithm itself is simple and only involves a few matrix multiplications
and Fourier transforms, which are easy to implement and efficient. There is no
need for supercells.
The cutoff ranges of the interactions are controlled by the $\vk$ and $\qp$ grids.
This real-space truncation is supported by the
``nearsightedness''~\cite{kohn1959analytic, prodan2005nearsightedness} of
electron--electron and electron--phonon interactions in the Wannier
representation. The corresponding real-space couplings decay with distance and
can therefore be truncated beyond a finite cutoff without strongly affecting the
results. In practice, this approximation is well satisfied for localized magnetic
exchange and for the short-range component of the EPC~\cite{PhysRevLett.98.047005,
PhysRevB.76.165108}.

In polar insulators, however, longitudinal optical phonons generate a long-range
dipolar Fr\"ohlich electron--phonon interaction~\cite{frohlich1954electrons} that
decays slowly in real space and is divergent at $\qp\to 0$. Modern EPC
implementations usually handle this by separating the EPC
into a short-range part, which is Wannier-interpolated, and a long-range
analytical part that describes the macroscopic dipole field in reciprocal
space~\cite{verdi2015frohlich, ponce2016epw}. In the
mixed-space approach implemented in \texttt{EPW}, the long-range part is
subtracted on the coarse grid and added back analytically on the fine grid.
Recent developments have further extended this separation to include dynamical
quadrupoles and higher-order multipoles~\cite{PhysRevLett.125.136601,
PhysRevB.102.094308}, as well as specific treatments for two-dimensional systems
where screening is qualitatively different~\cite{sohier2016two,
PhysRevB.103.115403, ponce2023longrange}.

We compute the SLC with a series of $\vk$-point and $\qp$-point meshes in the EPC calculation. With $n_{\vk}=3$ and $n_{\qp}=3$, the values are far from the finite-difference results and also from PT results with denser $\vk$ and $\qp$ meshes. For $n_{\qp}=5$, the $n_{\vk}=10$ and $n_{\vk}=15$ values converge and are also close to the FD results.
 Our present SLC formalism implicitly inherits this short-/long-range partition: the localized Wannier basis captures the short-range contribution, while any explicit long-range dipolar or multipolar correction must be added consistently at the EPC level. As a result, the locality assumption and the chosen real-space cutoffs should be carefully validated when applying the method to polar materials or when long-wavelength phonons are important.

The proposed method generates the exchange parameters $J_{IJ}^\mathrm{iso}$ and their
gradients $dJ_{IJ}^\mathrm{iso}/d\tauu$ in real space, corresponding to local spin
configurations and atomic displacements. These quantities are sufficient to build
the full momentum-dependent coupling between collective excitations, namely
magnons and phonons. One route is analogous to Wannier interpolation for EPC:
just as the electronic Hamiltonian is transformed from the localized Wannier
basis to Bloch states to reconstruct band structures, the magnetic Hamiltonian
can be transformed to momentum space. Because the $J$ parameters decay rapidly
with distance, the full magnon band structure $\omega(\vec{k})$ can be computed
efficiently using LSWT. The
local derivatives with respect to $\tauu$ can then be projected onto phonon
eigenvectors, which constructs the MPC at any point in the Brillouin zone.


The orbital resolution illustrated above is another useful feature of the
method, because it connects the exchange and SLC parameters to the electronic
channels that generate them.

While we demonstrate the method for a collinear spin system without SOC, the
generalization of the SLC to the DMI and anisotropic exchange is also
proposed here. Once the DFPT method and WF interpolation methods are implemented
for these cases, the additional terms can be calculated, which will enable the simulation
of many interesting properties. For example, the Gilbert damping in
atomistic spin dynamics is partly due to thermal noise from the phonons. With
the coupled model this phenomenon can be simulated explicitly. The spectra of magnons and
phonons can also be calculated at finite temperature by considering their mutual
interaction.

The approximations used in the derivation also define the limitations of the
method. The first limitation is the locality assumption discussed above,
especially for polar materials where long-range dipolar or multipolar EPC terms
may be important. The second limitation is that the spin--spin interaction is
described by a classical Heisenberg model, which neglects possible
non-Heisenberg terms and relies on a rigid-spin picture. In this approximation,
the magnitude of each local moment is fixed and only its orientation changes,
while the onsite response $\delta\mat{p}_I$ in Eq.~\eqref{eq:dAdvqp} is treated
as a small correction. In more itinerant systems, or close to a magnetic
instability, lattice distortions can modify not only the exchange couplings
between sites but also the size of the local moments themselves, corresponding
to longitudinal spin fluctuations. In such cases, the rigid-spin approximation
may become inaccurate, and a more general spin--lattice model in which both the
direction and magnitude of the spins are dynamical variables would be more
appropriate.

A related limitation is the neglect of higher-order spin interactions. The LKAG
method has recently been extended to include higher-order terms, such as
biquadratic, three-spin, and four-spin interactions~\cite{LKAGanharmonic,
LKAGanharmonic2}. These extensions could be adapted to the present formalism
fairly straightforwardly. Finally, the Heisenberg model also fails for strongly
itinerant magnetic moments, which are beyond the scope of the current method.

In some systems, second- or higher-order derivatives of the exchange with
respect to atomic displacements may be important. The current method can be
extended by including terms with two lattice perturbations, as illustrated for
higher-order SLC in Appendix~\ref{app:higher-order}.


%

\section{Conclusion}
\label{sect:conclusion}
To summarize, we have presented a method that combines perturbations to the spin
and lattice degrees of freedom to compute the SLC parameters. Our approach uses
the localized Wannier-function representation of the electronic Hamiltonian and
the electron--phonon coupling matrix as input, both of which can be obtained directly
from DFPT primitive-cell calculations with minimal extra computational cost.

The method is computationally efficient: for SrMnO$_3$, it reduces the cost by two orders of magnitude compared to the finite-difference approach, and this advantage grows
with the number of displacement perturbations. Moreover, by working directly in
the primitive cell with a dense $\bf q$-point grid, the perturbative approach
effectively bypasses finite-size supercell errors, making it potentially more
accurate than supercell-based finite-difference methods.
The method can be extended to other
types of spin interactions and higher-order phonon terms. Furthermore, it can
provide detailed information on the electronic and orbital origin of the SLC. We
illustrated this method with the example of SrMnO$_3$. The resulting real-space
parameters provide a direct route to parameterizing spin--lattice Hamiltonians
for finite-temperature spin--lattice dynamics using standard perturbative workflows.

\begin{acknowledgments}
The authors acknowledge initial discussions with Nicole Helbig on the spin-phonon coupling formalism and its symmetry properties.
MJV, EB and XH acknowledge the ARC AIMED project (G.A. 15/19-09)
funded by the Communaut\'e Fran\c{c}aise de Belgique.
MJV acknowledges funding by the Dutch Gravitation program
“Materials for the Quantum Age” (QuMat, reg number 024.005.006), financed by the Dutch Ministry of Education, Culture and Science (OCW).
EB acknowledges the SHAPEme project No.
560400077525 from the Belgian Excellence of Science (EOS)
program.
X. He acknowledges support from the Fonds de la Recherche Scientifique (FNRS) through the PDR project PROMOSPAN (Grant No. T.0107.20).
S. P. is a Research Associate of the Fonds de la Recherche Scientifique - FNRS.
This work was supported by the Fonds de la Recherche Scientifique - FNRS under Grants number T.0183.23 (PDR) and  T.W011.23 (PDR-WEAVE). 
This publication was supported by the Walloon Region in the strategic axe FRFS-WEL-T.
Computational resources have been provided by the Consortium des Équipements de Calcul Intensif (CÉCI), funded by the Fonds de la Recherche Scientifique de Belgique (F.R.S.-FNRS) under Grant No. 2.5020.11 and by the Walloon Region. 
We benefited from Lucia, the Tier-1 supercomputer of the Walloon Region, funded by the Walloon Region under the grant agreement n°1910247, and from EuroHPC-JU award EHPC-EXT-2023E02-050 on MareNostrum 5 at Barcelona Supercomputing Center, Spain.
\end{acknowledgments}

\section*{Data Availability}
The \texttt{TB2J} code, including the spin--lattice coupling feature developed in this work, is available at \url{https://github.com/mailhexu/TB2J}. The \texttt{Quantum ESPRESSO} and \texttt{EPW} codes are available at \url{https://www.quantum-espresso.org/} and \url{https://epw-code.org/}, respectively. The SrMnO$_3$ example data for computing the SLC parameters are openly available on Materials Cloud~\cite{SMOdata}.

\appendix

\section{Higher-order spin-phonon coupling \label{app:higher-order}}

For the pure spin exchange interaction, the Green's function involves two spin-rotation
perturbations (denoted by $\Delta$); thus, schematically the exchange arises from a $G \Delta G \Delta G$ type process, where $G$ is the Green's function. By adding one lattice perturbation ($g$), the SLC parameters arise from the combined perturbation, $G \Delta G g G \Delta G$. The extension to multiple spin and lattice distortions can be achieved in a similar way. Within the same linear-response approximation used for the single-displacement case [Eq.~\eqref{eq:greenpert}], the mixed second derivative of the Green's function with respect to two atomic displacements $u$ and $v$ is obtained by inserting two lattice vertices.

If we focus on the Green's-function variations and keep $\Delta_I$ and $\Delta_J$ fixed, a compact schematic form is
\begin{equation*}
  \mathcal{T}^{(2)}_{u,v} = \Delta_I \left( \delta_{\tau_u\tau_v}^2 G_{IJ}^{\uparrow} \right) \Delta_J G_{JI}^{(0),\downarrow} + \Delta_I G_{IJ}^{(0),\uparrow} \Delta_J \left( \delta_{\tau_u\tau_v}^2 G_{JI}^{\downarrow} \right) ,
\end{equation*}
where the second-order Green's-function variation can be written as a sum of ``Fan--Migdal'' and ``Debye--Waller'' terms,
\begin{equation*}
  \begin{aligned}
  \delta_{\tau_u\tau_v}^2 G = &\underbrace{G^{(0)} g_u G^{(0)} g_v G^{(0)} + G^{(0)} g_v G^{(0)} g_u G^{(0)}}_{\text{Fan--Migdal terms}} \\
  &+ \underbrace{G^{(0)} g_{uv}^{(2)} G^{(0)}}_{\text{Debye--Waller term}} ,
  \end{aligned}
\end{equation*}
with $g_u = \partial V / \partial \tau_u$ the first-order electron--phonon vertex and $g_{uv}^{(2)} = \partial^2 V / \partial \tau_u \partial \tau_v$ the second-order vertex. In addition there are cross terms where both propagators are differentiated once,
\begin{equation*}
\begin{aligned}
  \mathcal{T}^{\text{cross}}_{u,v} = \Delta_I \left( \delta_{\tau_u} G_{IJ}^{\uparrow} \right) \Delta_J \left( \delta_{\tau_v} G_{JI}^{\downarrow} \right) +\\ \Delta_I \left( \delta_{\tau_v} G_{IJ}^{\uparrow} \right) \Delta_J \left( \delta_{\tau_u} G_{JI}^{\downarrow} \right) ,
\end{aligned}
\end{equation*}
which, using $\delta_{\tau_u} G^{\uparrow,\downarrow} = (G^{(0),\uparrow,\downarrow} g_u G^{(0),\uparrow,\downarrow})$, can be written schematically as
\begin{equation*}
  \mathcal{T}^{\text{cross}}_{u,v} = \Delta_I \left( G_{IJ}^{(0),\uparrow} g_u G_{IJ}^{(0),\uparrow} \right) \Delta_J \left( G_{JI}^{(0),\downarrow} g_v G_{JI}^{(0),\downarrow} \right) + (u \leftrightarrow v) .
\end{equation*}
A further class of terms comes from derivatives of the magnetic potentials themselves. Second-order derivatives of the onsite exchange splittings give a ``single-site Hessian'' contribution
\begin{equation*}
  \mathcal{T}^{(\Delta\Delta)}_{u,v} = (\delta_{\tau_u\tau_v}^2 \Delta_I) \, G_{IJ}^{\uparrow} \, \Delta_J \, G_{JI}^{\downarrow} + \Delta_I \, G_{IJ}^{\uparrow} \, (\delta_{\tau_u\tau_v}^2 \Delta_J) \, G_{JI}^{\downarrow} ,
\end{equation*}
where $\delta_{\tau_u\tau_v}^2 \Delta_I = \partial^2 \Delta_I / (\partial \tau_u \, \partial \tau_v)$ measures the curvature of the local exchange splitting with respect to lattice displacements. There are also response--response cross terms,
\begin{equation*}
  \begin{aligned}
  \mathcal{T}^{(\delta\Delta\text{-cross})}_{u,v} = &(\delta_{\tau_u} \Delta_I) \, G_{IJ}^{\uparrow} \, (\delta_{\tau_v} \Delta_J) \, G_{JI}^{\downarrow} \\
  &+ (\delta_{\tau_v} \Delta_I) \, G_{IJ}^{\uparrow} \, (\delta_{\tau_u} \Delta_J) \, G_{JI}^{\downarrow} ,
  \end{aligned}
\end{equation*}
and a larger set of mixed terms where one derivative acts on $\Delta$ and the other on $G$,
\begin{equation*}
\begin{aligned}
  \mathcal{T}^{(\text{mixed})}_{u,v} = \;&
  (\delta_{\tau_u} \Delta_I) (\delta_{\tau_v} G_{IJ}^{\uparrow}) \Delta_J G_{JI}^{\downarrow}
  + (\delta_{\tau_v} \Delta_I) (\delta_{\tau_u} G_{IJ}^{\uparrow}) \Delta_J G_{JI}^{\downarrow} \\
  & + \Delta_I (\delta_{\tau_u} G_{IJ}^{\uparrow}) (\delta_{\tau_v} \Delta_J) G_{JI}^{\downarrow}
  + \Delta_I (\delta_{\tau_v} G_{IJ}^{\uparrow}) (\delta_{\tau_u} \Delta_J) G_{JI}^{\downarrow} \\
  & + (\delta_{\tau_u} \Delta_I) G_{IJ}^{\uparrow} \Delta_J (\delta_{\tau_v} G_{JI}^{\downarrow})
  + (\delta_{\tau_v} \Delta_I) G_{IJ}^{\uparrow} \Delta_J (\delta_{\tau_u} G_{JI}^{\downarrow}) \\
  & + \Delta_I G_{IJ}^{\uparrow} (\delta_{\tau_u} \Delta_J) (\delta_{\tau_v} G_{JI}^{\downarrow})
  + \Delta_I G_{IJ}^{\uparrow} (\delta_{\tau_v} \Delta_J) (\delta_{\tau_u} G_{JI}^{\downarrow}) .
\end{aligned}
\end{equation*}
Collecting all these pieces, a compact schematic expression for the full second-order coupling reads
\begin{equation*}
  \begin{aligned}
    \frac{\partial^2 J_{IJ}}{\partial \tau_u \, \partial \tau_v} = &\; \frac{1}{4\pi} \, \Im \int_{-\infty}^{E_F} d\epsilon \, \Tr_{L} \Big[ \mathcal{T}^{(2)}_{u,v} + \mathcal{T}^{\text{cross}}_{u,v} \\
    & + \mathcal{T}^{(\Delta\Delta)}_{u,v} + \mathcal{T}^{(\delta\Delta\text{-cross})}_{u,v} + \mathcal{T}^{(\text{mixed})}_{u,v} \Big] .
  \end{aligned}
\end{equation*}

Most of the terms are already available in the process of computing the single-atomic-displacement SLC.
However, terms including the second-order derivative of the electronic potential with respect to atomic displacements, such as for the Debye--Waller term in the Allen-Heine-Cardona approach~\cite{allen1976theory}, require more care. The calculation of these parameters can be performed with DFPT and they are available as output in the \texttt{Abinit} package.
Higher derivatives will be computationally costly.

\bibliography{mybib}

\end{document}